# Reduction of Redundant Rules in Association Rule Mining-Based Bug Assignment

Meera Sharma*,§, Abhishek Tandon†,¶,
Madhu Kumari‡, and V. B. Singh‡,**

*Swami Shrddhanand College
University of Delhi, Delhi, India

†SSCBS, University of Delhi, Delhi, India

‡Delhi College of Arts and Commerce
University of Delhi, Delhi, India

§meerakaushik@gmail.com
¶abhishektandon86@gmail.com
mesra.madhu@gmail.com
**vbsinghdcacdu@gmail.com

Bug triaging is a process to decide what to do with newly coming bug reports. In this paper, we have mined association rules for the prediction of bug assignee of a newly reported bug using different bug attributes, namely, severity, priority, component and operating system. To deal with the problem of large data sets, we have taken subsets of data set by dividing the large data set using *K*-means clustering algorithm. We have used an Apriori algorithm in MATLAB to generate association rules. We have extracted the association rules for top 5 assignees in each cluster. The proposed method has been empirically validated on 14,696 bug reports of Mozilla open source software project, namely, Seamonkey, Firefox and Bugzilla. In our approach, we observe that taking on these attributes (severity, priority, component and operating system) as antecedents, essential rules are more than redundant rules, whereas in [M. Sharma and V. B. Singh, Clustering-based association rule mining for bug assignee prediction, *Int. J. Business Intell. Data Mining* 11(2) (2017) 130–150.] essential rules are less than redundant rules in every cluster. The proposed method provides an improvement over the existing tech-niques for bug assignment problem.

## 1. Introduction

An essential aspect of software development projects demands an efficient bug tracking system. Bug tracking systems are of huge importance in open source soft-ware development. The essence of bug tracking systems is dispersing information related to bug to software development team, which is distributed worldwide.[1] The efficiency of bug tracking systems is even more important because worldwide distribution team members might not have ever met or seen each other. Hence, looking at problem of worldwide of "team coordination" becomes the buzz word of bug tracking systems. Hence, bug tracking system plays an extremely important role in tracking bug reports and coordinating among team distributed worldwide among possibly unknown team members. Thus, the efficiency of bug tracking systems may be understood broadly as an optimized way of tracking bug reports generated in a given time frame (usually 1 day) and coordination. There exists lots of scope in terms of the coordination of bug tracking systems, which explores a new dimension of research. One would rather aim to generate complete automatic coordination with minimum, preferably zero manual interference. To illustrate the efficiency of bug tracking systems in an explicit manner, let us understand maintenance issues related to "Bugzilla". Bugzilla popularity could be measured as it was used by projects such as Eclipse,[12] KDE and Gnome.[10] Each of the applications was get-ting hundreds of bug reports. These projects were expected to assign bugs to the relevant expert which might be able to solve in minimum time; hence optimiza-tion of the bug problems was obtained. The process which distributes the bug to a given software developer team member is known as "bug assignment".[3,9,11] The bug assignment is not only a complicated process, but also time-consuming. The bug triaging process is labor-intensive, time-consuming and worst fault-prone because of the judgment of a person which may vary emotionally, if anybody do this manu-ally.[4] Thus, this area of bug assignment is really important to automate. It is also very difficult to manually remember the expertise of various software developers. Manual bug assignment process hurts even more since ever increasing bug reports may aggravate the issue of new bug if not assigned to the right software developer. The reported bug must be triaged; bug triaging determines meaningful enhance-ment to new bug reported and assigned to an appropriate developed for further handling. Software bugs are unavoidable, and fixing of bug is must but costly dur-ing the development of an open source software project. To improve efficiency and to reduce costs, a person indicates a bug report that fixes a number of technical issues to the appropriate fixer. This process is known as a bug triaging.[5] Bug triag-ing means when the newly coming bug reports are submitted, it must be assigned to the right fixer to fix it.

**Corresponding author

A bug is defined by several attributes.[20] Among all the attributes, we have used five quantified attributes, namely, severity, priority, component, operating system and assignee. Severity indicates how severe a bug is. There are seven levels of sever-ity from 1 to 7, namely, blocker, critical, major, normal, minor, trivial and enhance-ment. Level 1 is the blocker which is more severe and level 7 is the enhancement which is least severe.[15–18] Bug priority describes the significance and state in which a bug should be fixed. This field is used by the assignee to prioritize his or her bugs. The available priorities range from P1 (most important) to P5 (least important).[19] Components are subdivisions of a product. Operating system consists of diff erent types of operating system in respect of which bug was reported. The bug assignee is a person to whom a bug is assigned to fix. In this paper, we have collected 14,696 bug reports of Seamonkey, Firefox and Bugzilla products of the Mozilla open source project. Instead of taking a large data set, we have divided the data sets into five clusters using $K$ -means clustering algorithm. In each cluster, we have applied asso-ciation rule mining for top 5 assignee using Apriori algorithm based on bug severity, priority, component and operating system.

In the paper,[13] the authors have first discussed association rule mining method. A rule-based machine learning method for determining the patterns of co-existence of the attributes in a database is called association rule mining. An association rule mining method is defined by two disjoint item sets in the form $X \Rightarrow Y$, where $X \cap Y = \phi$. Every rule is combined by two different sets of items known as the antecedent and consequent. Here $X$ refers as an antecedent of the rule and $Y$ refers as a consequence of the rule.[14,21]

The importance, strength and certainty of an association rule mining are deter-mined by support and confidence. Frequency of a rule in a given data set may be expressed as support, where confidence may be determined by how many times item $Y$ is present of transaction of item $X$. Association rule mining based on significance and certainty produces large number of essential and redundant rules. Redundant rules are major issues in association rule mining algorithm. Generally, the number of redundant rules is larger than that of the essential rules, but in our study, we have obtained as essential rules are larger than the redundant rules when the data size is less and the number of redundant rules is larger than that of the essential rules when the data size is large. In a study,[14] the authors state that a set of asso-ciation rules $R$ which is obtained from a set of frequent item sets $I$, where each element $X \in R$, satisfy both *significance* and *certainty* thresholds. A rule $X$ in $R$ is said to be *redundant* if and only if a rule or a set of rules $S$ where $S \in R$ possess the same intrinsic meaning of $X$. We have identified and eliminated redundant rules from our results by using the techniques described by Ashrafi *et al.*[14]

Several associative classification studies for various applications have been con-ducted.[22–29] We organized the rest of the paper as follows. Section 2 presents the description of data sets and its preprocessing. Results have been presented in Sec. 3. Related works have been described in Sec. 4. Section 5 presents threats to validity and finally we conclude the paper in Sec. 6.

## 2. Description of Data Sets and Data Preprocessing

In our study, we have collected 14,696 bug reports of the open source software products (Mozilla), namely, Seamonkey, Firefox and Bugzilla. We have conducted empirical experiment on bug report for resolution “fixed” and status “verified”, “resolved” and “closed” because these bug reports consist of meaningful information for the experiment. Table 1 shows the number of bug reports and the period of time during which the bug reports were collected in each product.

Table 1. Number of bug reports in each product.

| Product | Total no. of bugs | Observation period |
|---|---|---|
| Seamonkey | 6,613 | Apr. 1998–Aug. 2016 |
| Firefox | 6,148 | Apr. 2001–Oct. 2016 |
| Bugzilla | 1,935 | Sept. 1994–Aug. 2016 |

We have taken five quantified attributes, namely, severity, priority, component, operating system and assignee for analysis. We have used an Apriori algorithm to extract the association rules in MATLAB software using ARMADA tool to predict bug assignee using bug component, operating system, severity and priority.

The following steps have been conducted for our study.

*Data extraction*

- Downloaded 14,696 bug reports of three products: Seamonkey, Firefox and Bugzilla of Mozilla open source project.
- Saved the downloaded file in excel format for further processing.

*Data preparation*

- Assign numeric value 1 to 7 to the severity attribute and 1 to 5 to the priority attribute.
- Assign numeric value to the component, operating system and assignee attribute.

*Modeling* (*K-means clustering and association rule mining*)

- We have divided each large data set using $K$ -means clustering algorithm in Rapid Miner into five clusters and then apply association rule mining on every cluster using bug severity, priority, component and operating system to predict the bug assignee.
- We have extracted rules using Apiori Algorithm with ARMADA (Association Rule Miner And Deduction Analysis) tools in MATLAB. Minimum support count 3 and minimum confidence 10% are taken to extract the association rules in this paper.

*Testing and validation*

- We validate the result of association rule mining based on performance measures, namely, the significance (support count) and certainty (confidence).

## 3. Results and Discussion

In this paper, we have applied $K$ -means clustering algorithm in Rapid Miner tool to divide the data set into five clusters as the data set is large. After that we have mined association rule on every cluster to predict assignee with severity, priority, compo-nent and operating system as antecedents. To extract the rules, we applied an Apri-ori algorithm using ARMADA tool in MATLAB software with support 3 as count

Table 2. Association rules in cluster 1 for Seamonkey.

1. Severity {Normal} Λ Priority {P3} Λ Os {Linux} Λ Component{Build Config} ⇒ Assignee {Jon Granrose} @ (9,52.94%)
2. Severity {Major} Λ Priority {P3} Λ Os {All} Λ Component{MailNews: Message Display} ⇒ Assignee {Varada} @ (3,75%)
3. Severity {Normal} Λ Priority {P3} Λ Os {All} Λ Component{MailNews: Backend} ⇒ Assignee {Diane Sun} @ (4,66.67%)
4. Severity {Normal} Λ Priority {P3} Λ Os {Linux} Λ Component{General} ⇒ Assignee {Akkana Peck} @ (3,60%)
5. Severity {Normal} Λ Priority {P3} Λ Os {All} Λ Component{MailNews: Message Display} ⇒ Assignee {Suresh } @ (13,27.66%).

and confidence 10%. In every cluster, we have considered top 5 assignees based on the number of bugs assigned to them and extract the rules for those assignees. As a result, we have obtained 1-antecedent, 2-antecedent, 3-antecedent and 4-antecedent rules for predicting assignee. We get more than 100 rules in each cluster. For this reason, we present one 4-antecedent rule for every assignee in each cluster.

In Table 2, we have shown five 4-antecedent association rules based on the high-est certainty for severity, priority, component and operating system as antecedents and assignee as consequent in cluster 1 for Seamonkey product. Here we have listed the results of one cluster due to limitations of pages.

The above rules reveal that *Jon Granrose* is an assignee having severity *Normal*, priority P3, Operating System *Linux* and Component *Build Config* with a support count of 9 and confidence of 52.94%. The second rule shows that for severity Major, priority P3 Operating System *All* and Component *MailNews*: Message Display, *Varada* is an assignee with a support count of 3 and confidence of 75%. The third rule shows that *Diane Sun* is an assignee to fix the bug having severity *Normal*, Priority P3, Operating System *All* and Component *MailNews*: *Backend* with a support count of 4 and confidence of 66.67%. The fourth rule shows that for severity *Normal*, priority P3, operating system *Linux* and component G*eneral*, *Akkana Peck* is an assignee with a support count of 3 and confidence of 60%. The fifth rule shows that for severity *Normal*, priority P3, operating system *All* and Component *MailNews*: *Message Display*, *Suresh* is an assignee to fix the bug with a support count of 13 and confidence of 27.66%.

In Table 3, we have shown five 4-antecedents association rules based on the high-est certainty for severity, priority, component and operating system as antecedents and assignee as consequent in cluster 1 for Firefox product. Here, we have listed the results of one cluster due to limitations of pages.

The first rule shows that *Blake Ross* is an assignee to fix the bug having severity *Normal*, priority P4, operating system *All* and component *General* with a support count of 4 and confidence of 66.67%. In the second rule, *Kit Cambridge* is an assignee for severity *Normal*, priority P1, operating system *Unspecified* and component *Sync* with a support count of 7 and confidence of 77.78%. The third rule reveals that

Table 3. Association rules in cluster 1 for Firefox.

1. Severity {Normal} Λ Priority {P4} Λ Os {All} Λ Component{General} ⇒ Assignee {Blake Ross} @ (4,66.67%)
2. Severity {Normal} Λ Priority {P1} Λ Os {Unspecified} Λ Component{Sync} ⇒ Assignee {Kit Cambridge} @ (7,77.78%)
3. Severity {Normal} Λ Priority {P3} Λ Os {Unspecified} Λ Component{Developer Tools: Debugger} ⇒ Assignee {Jason Laster} @ (7,100%)
4. Priority {P2} Λ Os {All} Λ Component{Bookmarks & History} ⇒ Assignee {Pierre Chanial} @ (3,50%)
5. Priority {P3} Λ Os {All} Λ Component{Preferences} ⇒ Assignee {Steffen Wilberg} @ (3,60%).

*Jason Laster* is an assignee with a support count of 7 and confidence of 100% having severity *Normal*, priority P3, operating System *Unspecified* and Component *Developer Tools*: *Debugger*. The fourth rule is a three antecedent rule, which shows that *Pierre Chanial* is an assignee to fix the bug having priority P2, operating system *All* and Component *Bookmarks & History* with a support count of 3 and confidence of 50%. The fifth rule is a three antecedent rule, which reveals that the assignee *Steffen Wilberg* is an assignee with a support count of 3 and confidence of 60% having priority P3, operating system *All* and component *Preferences*.

In Table 4, we have shown five association rules based on the highest certainty for severity, priority, component and operating system as antecedents and assignee as consequent in cluster 1 for Bugzilla product. Here we have listed the results of one cluster due to limitations of pages.

The first rule is a three antecedent rule, which reveals that the assignee *Dan Mosedale* can be assigned a bug having severity *Normal*, priority P3 and Component *Bugzilla-General* with a support count of 3 and a confidence of 30%. The second rule shows that for severity level *Normal*, priority P3 and component *Bugzilla-General*, the assignee *Christine Begle* can be assigned the bug with a support count of 3 and confidence of 30%. In the third rule, the assignee *Toms Baugis* can be assigned the bug having severity *Enhancement* and priority P3 with a support count of 3 and confidence of 16.67%. The fourth rule is two antecedents rule, which shows that *Adam Kennedy* is an assignee having severity *Trivial* and operating system *All* with

Table 4. Association rules in cluster 1 for Bugzilla.

1. Severity {Normal} Λ Priority {P3} Λ Component{Bugzilla-General} ⇒ Assignee {Dan Mosedale} @ (3,30%)
2. Severity {Normal} Λ Priority {P3} Λ Component{Bugzilla-General} ⇒ Assignee {Christine Begle} @ (3,30%)
3. Severity {Enhancement} Λ Priority {P3} Λ Os {All} ⇒ Assignee {Toms Baugis} @ (3,16.67%)
4. Severity {Trivial} Λ Os {All} ⇒ Assignee {Adam Kennedy} @ (3,23.08%)
5. Os {All} Λ Component{User Interface} ⇒ Assignee {Ben Goodger} @ (3,75%).

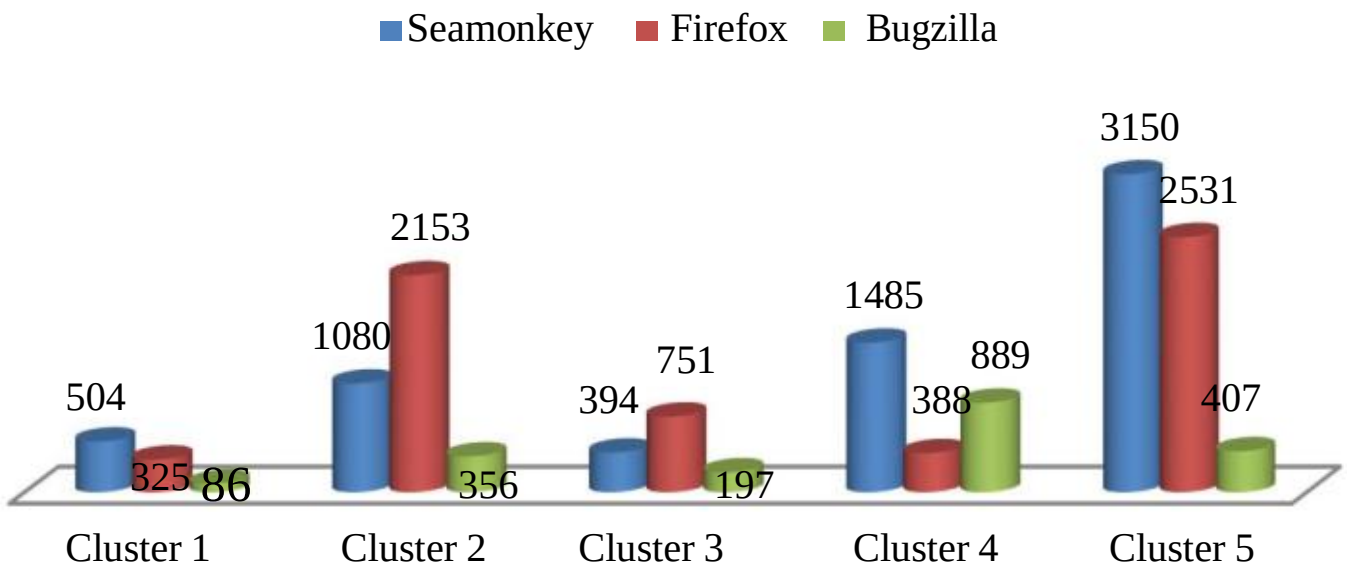

Fig. 1. Cluster-wise data distribution for Seamonkey, Firefox and Bugzilla.

a support count of 3 and a confidence of 23.08%. The fifth rule is two antecedent rule, which reveals that the assignee *Ben Goodger* can be assigned the bug having operating system *All* and Component *User Interface* with a support count of 3 and confidence of 75%.

In Fig. 1, we have shown the cluster wise data distribution for all the three products, namely, Seamonkey, Firefox and Bugzilla.

In Figs. 2–4, we have shown the cluster-wise essential association rules and redundant rules for all the three products, namely, Seamonkey, Firefox and Bugzilla. As a result, we have obtained that for large data size there are more redundant rules than essential rule and for small data size there are more essential rules than redundant rules.

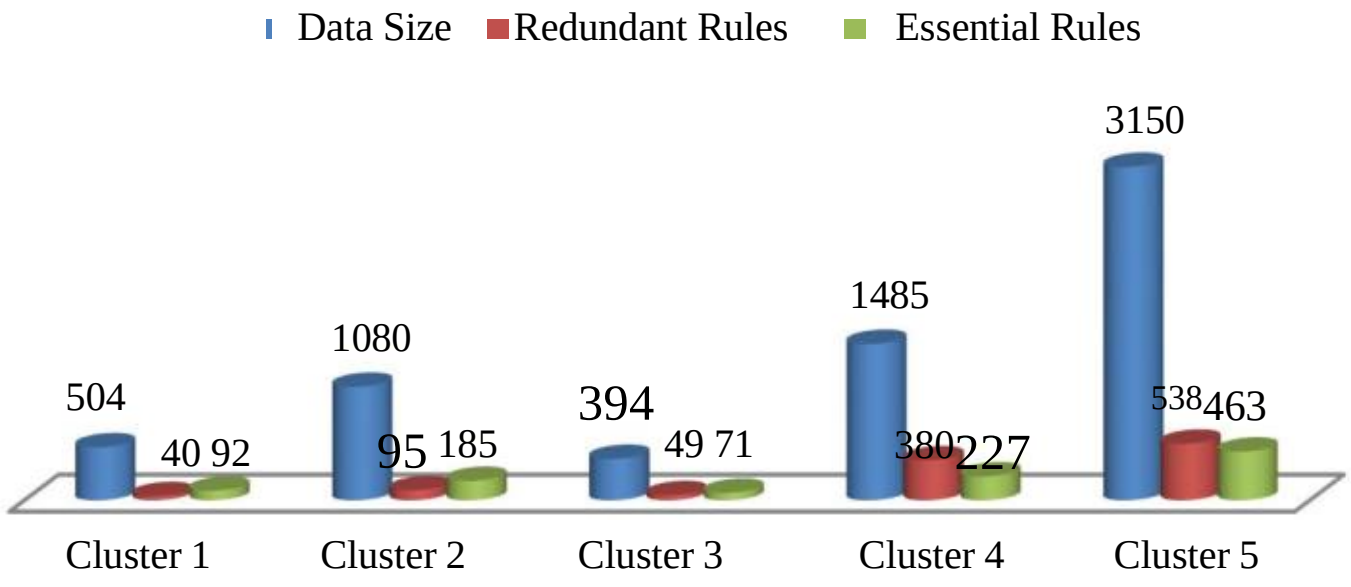

Fig. 2. Cluster wise redundant and essential rules distribution for Seamonkey.

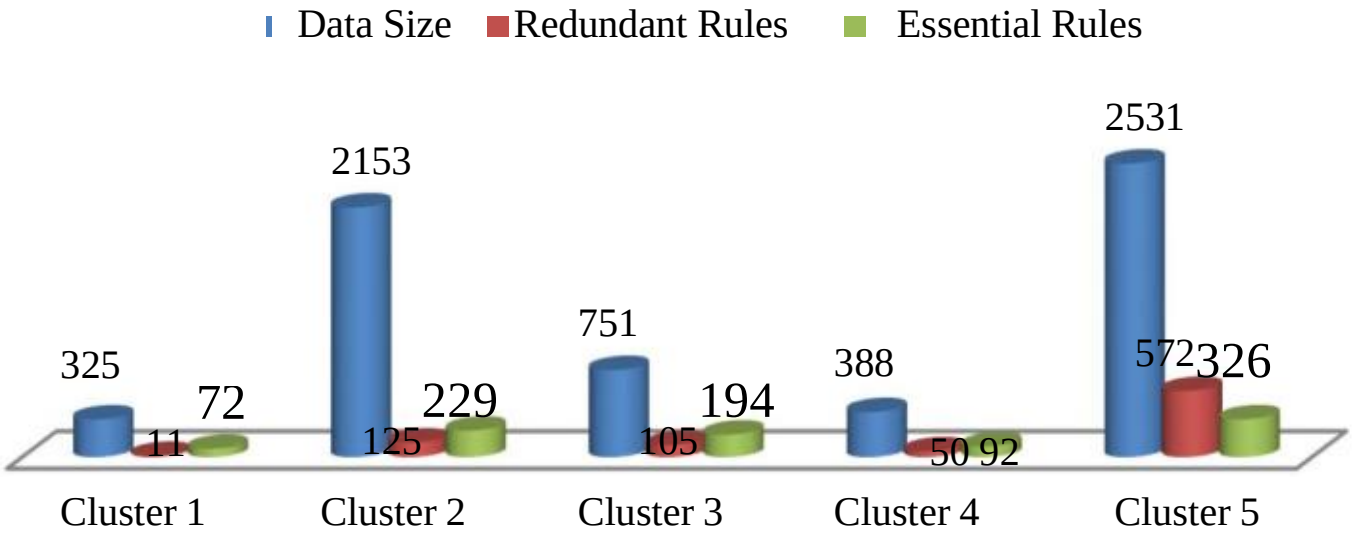

Fig. 3. Cluster-wise redundant and essential rules distribution for Firefox.

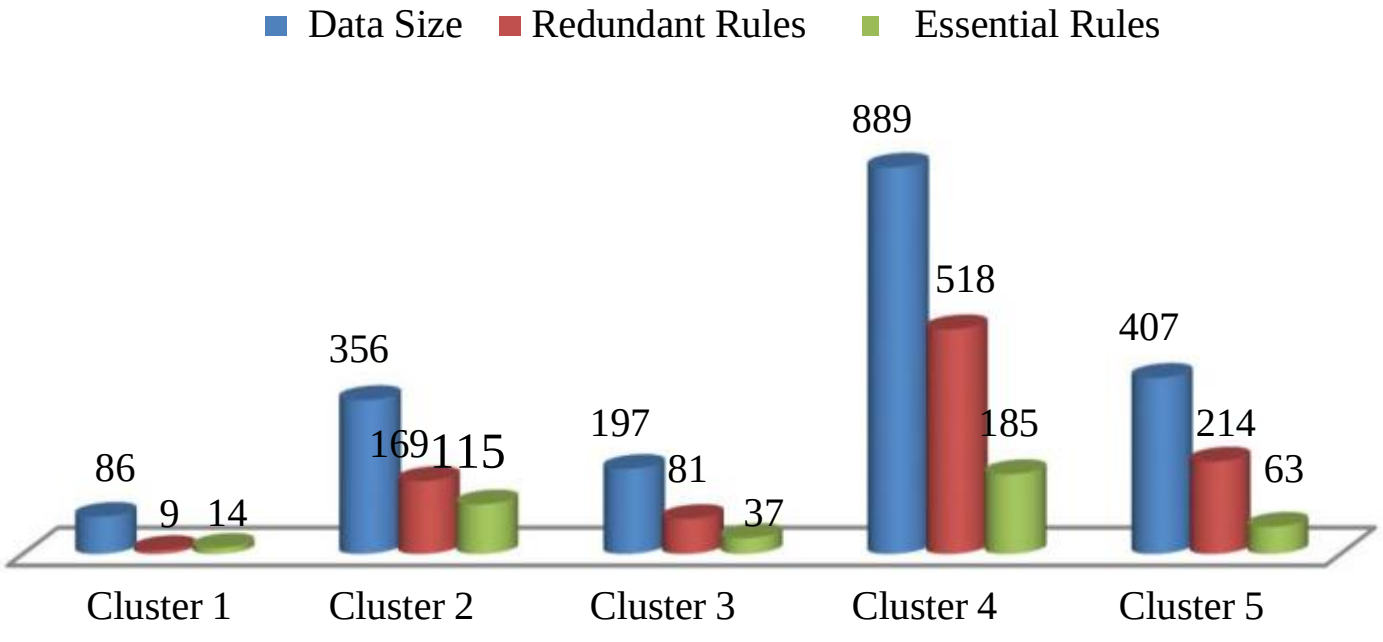

Fig. 4. Cluster-wise redundant and essential rules distribution for Bugzilla.

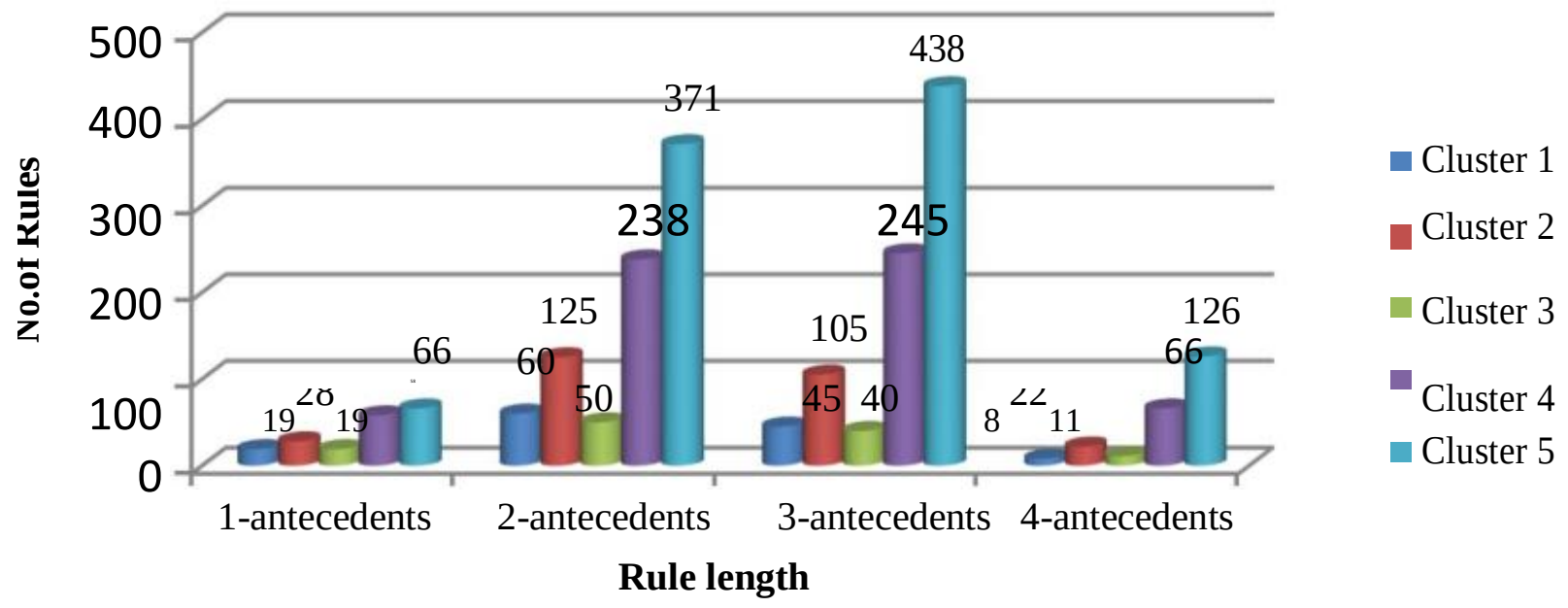

Fig. 5. Distribution of association rules by rule length in all clusters for Seamonkey.

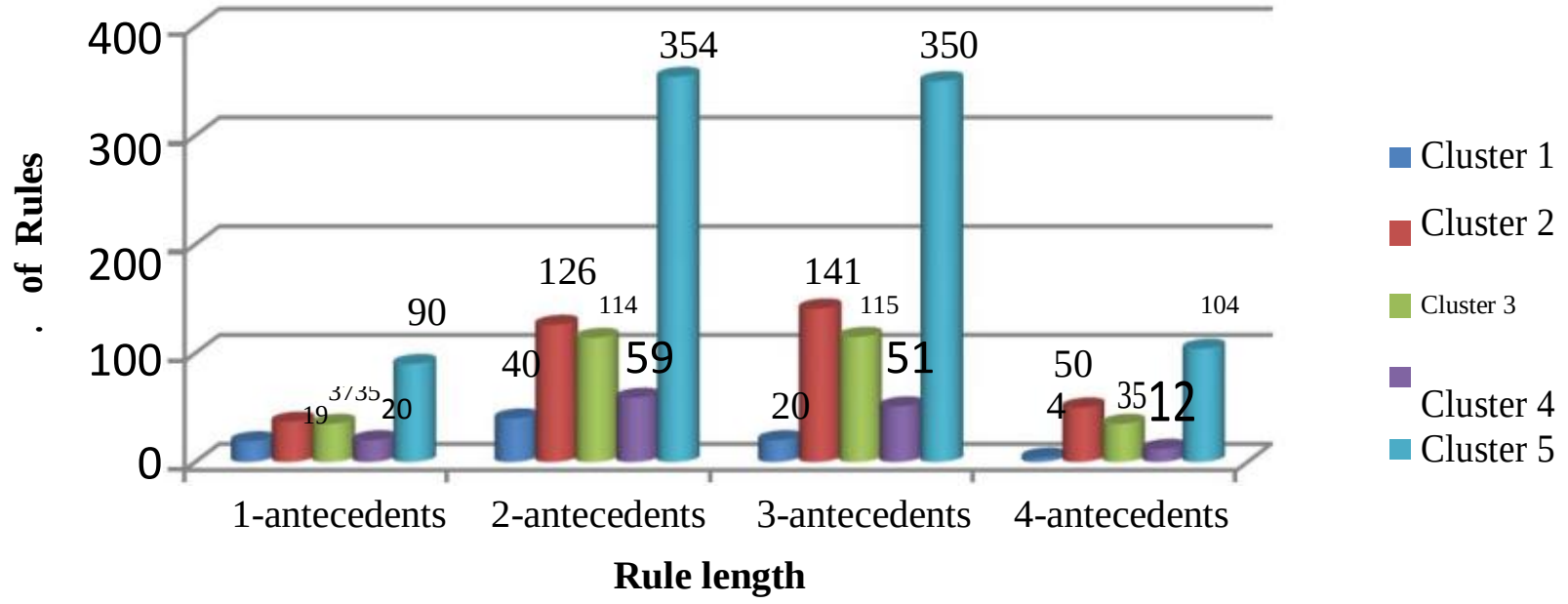

Fig. 6. Distribution of association rules by rule length in all clusters for Firefox.

So, if the data size increases, redundancy also increases.

In Figs. 5–7, we have shown the distribution of association rule according to their rule length in all the five clusters for all the three products.

Figures 5–7 show that there is a maximum number of association rules of two antecedents in all the five clusters for all the three products.

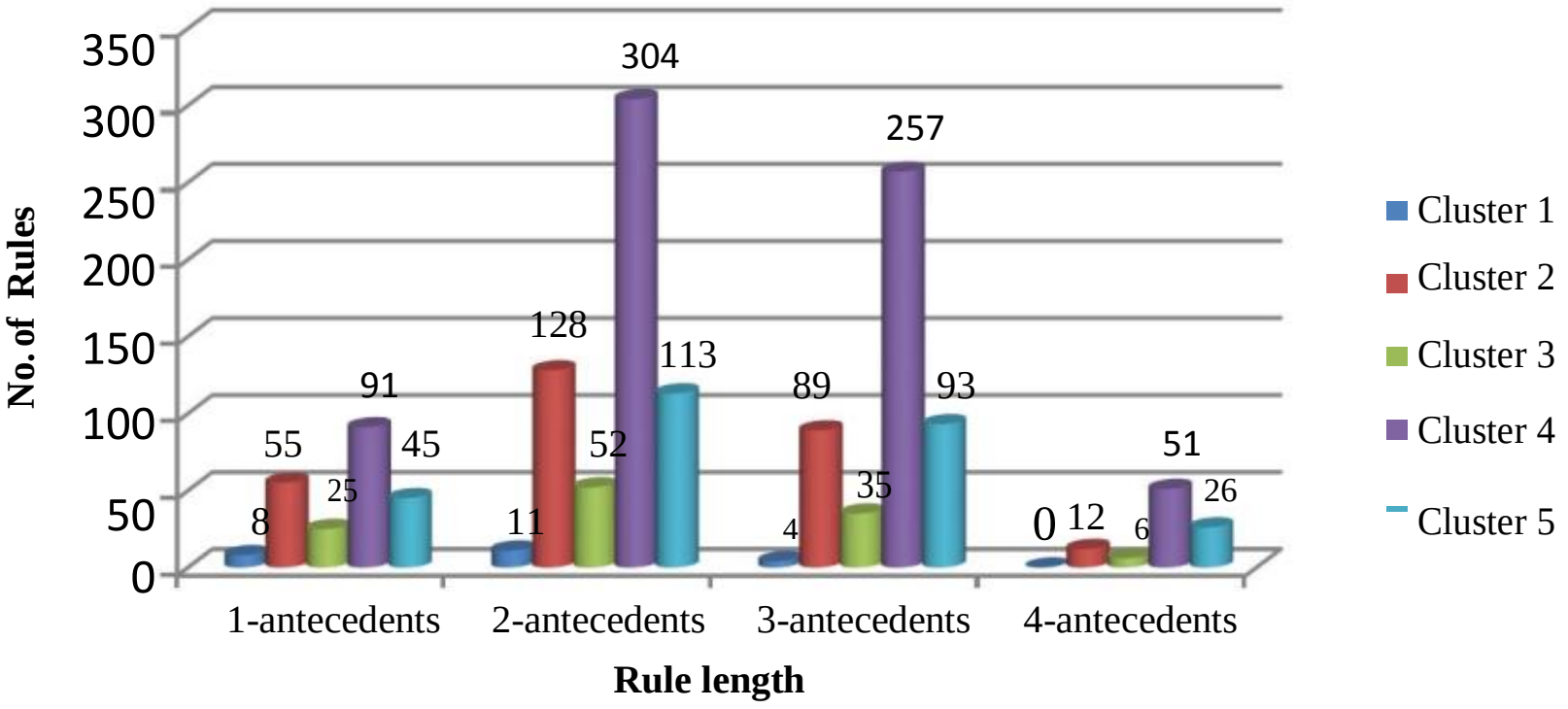

Fig. 7. Distribution of association rules by rule length in all clusters for Bugzilla.

## 4. Related Work

Bug assignee prediction is an active area of research. In paper,[1] the authors pre-dicted the name of developers to whom the bug can be assigned using machine learn-ing techniques. The empirical investigation was conducted on 15,859 bug reports of the Eclipse project. Performance measure accuracy has been used to measure the eff ectiveness of the built classifiers, and the accuracy was 30%. Later on, Anvik *et al.*[2] extended the work of Cubranic and Murphy[1] by applying a Na¨ıve Bayes classifier, nonlinear support vector machines and C4.5. The experiment was con-ducted on two data sets, namely, Eclipse and Firefox, and achieved the precision levels of 57% and 64%. The proposed method was validated using 5,200 bug reports of the Eclipse JDT project, and the performance measure was found to be 90.1% precision and 45.5% recall. In a study,[4] the authors proposed an optimization tech-nique using a Naive Bayes classifier with incremental learning and bug assign-ment graphs. The experiment was conducted on 856,259 bug reports of Mozilla and Eclipse projects and reduces the length of tossing graph by the prediction accuracy of 86.09%, and the prediction accuracy is improved by 10.78% compared with pre-vious approaches. In the study,[5] the authors have proposed a novel fuzzy set and cache-based approach for automatic bug triaging called Bugzie. Bugzie considers a fuzzy set software system which is associated with each technical term. Fuzzy set is used to represent the developer to fix the bugs relevant to each term. The value of the membership function in the fuzzy set is obtained from the bug reports that it has corrected and updated when the newly fixed bug reports are available. For a new bug report, to find the most potential fixers, based on the technical term, Bugzie combines the fuzzy sets and classifies the developers according to their value of the membership function. Bhattacharya and Neamtiu[6] demonstrate several approaches, namely, intrafold updates and refined classification. The authors achieved 83.62% bug assignment accuracy and reduce tossing path lengths. Further, a predictive model has been developed to allocate developers for bug reports using naive Bayes classification[7] and hence increases the efficiency of bug triage. The outline of the paper[8] is to introduce tossing graph model based on the Markov property in order to reduce tossing events, by up to 72%. The authors also achieved 23% prediction accuracy compared with other bug triaging approaches. In a study,[32] the authors used data reduction technique to reduce the data scale in bug triaging by including the representative values along with the statistical values of the bug data set. The authors demonstrate the work on an open source project Eclipse. The authors also achieved 96.5% of bug triaging accuracy, which is better than the existing work. The problem of data reduction has been proposed by Xuan *et al.*[33] for bug triage. The authors have used feature selection and instance selection to reduce data scale on the bug dimension and the word dimension. The performance of data reduction has been empirically investigated on 600,000 bug reports of two large open source projects, namely, Eclipse and Mozilla. Tian *et al.*[34] proposed a unified model that combines information from both developers' previous activities and suspicious pro-gram locations associated with a bug report in the form of similarity features. The approach has been validated on more than 11,000 bug reports from Eclipse JDT, Eclipse SWT and ArgoUML projects. Results showed that the model can outperform a location-based baseline[35] and an activity-based baseline.[36] A novel bug assignment approach, W8Prioritizer, based on bug parameter prioritization has been proposed by Goyal and Sardana.[37] The authors have extended the study for triaging of nonreproducible (NR) bugs. Whenever the developer faces any issue in reproducing a bug report, he/she marks the bug report as NR. However, certain portions of these bugs get reproduced and eventually fixed later. To predict the fixability of bug reports marked as NR, a prediction model, NRFixer, has been pro-posed. Association rule mining approach has been developed to predict the right developers.[30] The proposed association rule model has been validated using bug reports of Thunderbird, AddOnSDK and Bugzilla projects. Later on, Sharma and Singh[31] proposed two methods to apply association rule mining for bug assignee prediction. In the first method, the authors have used Apriori algorithm to predict the bug assignee based on bug's severity, priority and summary terms. In the second method, the authors divide the data set using $X$ -means clustering algorithm and apply association rule mining in each cluster. They validate their result on 1,695 bug reports of Thunderbird, Add-on SDK and Bugzilla. In this study, essential rules are less than redundant rules. In this paper, our approach is to reduce the redundant rules. So that the complexity of bug assignment based on different bug attributes can be reduced.

## 5. Threats to Validity

The following are the factors that affect the validity of our study:

*Construct validity*: We have considered bug attributes, namely, severity, priority, component, operating system and the bug assignee. These attributes are not based on any empirical verification.

*Internal validity*: In this paper, we have extracted the association rules to predict the developer of a newly reported bug using the bug attributes severity, priority, component and operating system. We have used ARMADA tool in MATLAB for Apriori algorithm. Other algorithms and more attributes can also be used.

*External validity*: We validate our approach on 14,696 bug reports of three products, namely, Seamonkey, Firefox and Bugzilla. We can extend our study for another open source and closed source software.

## 6. Conclusion

In this paper, we have applied the association rules to predict the bug assignee of a newly coming bug report based on the bug's severity, priority, component and oper-ating system. We have used an Apriori algorithm on the ARMADA tool in MAT-LAB to generate the association rules. We have divided the dataset into five clusters using $K$ -means clustering algorithm and extracted the rules for five assignees in each cluster. As a result, we obtained 1-Antecedent, 2-Antecedents and 4-Antecedent rules. We get the maximum number of association rules of 2-antecedents in all the five clusters for all the three products. We have also identified and eliminated the redundant rules. We have observed that in all clusters, if the data size increases, rules redundancy also increases and taking of attributes as severity, priority, com-ponent and operating system as antecedents to predict the assignee, redundancy of rules decreases when compared with the previous work.[31] In this paper, our con-tribution is to reduce the redundant rules by taking two new attributes, namely component and operating system.

## About the Authors

Meera Sharma is currently working as an Assistant Professor in the Department of Computer Science at Swami Shraddhanand College, University of Delhi, Delhi, India. She has published 18 papers in repute journals and conference proceedings. Her research interests include Mining Software Repositories and Open Source Soft-ware. Currently pursuing Ph.D. in Computer Science at Department of Computer Science, University of Delhi, Delhi, India.

Abhishek Tandon is currently working as an Assistant Professor in Shaheed Sukhdev College of Business Studies, University of Delhi, Delhi, India. He received his PhD degree in Software Reliability and Marketing from Department of Oper-ational Research, University of Delhi. He has published papers in the field of reli-ability modeling, optimization theory, forecasting and marketing research. He is a life member of the Society for Reliability Engineering, Quality and Operations Management (SREQOM).

Madhu Kumari is currently working as an Assistant Professor in the Department of Computer Science at Delhi College of Arts and Commerce, University of Delhi, Delhi, India. She has published nine papers in repute journals and conference pro-ceedings. Her research interests include Mining Software Repositories and Big Data Software Engineering. Currently pursuing Ph.D. in Computer Science at Depart-ment of Computer Science, University of Delhi, Delhi, India.

V. B. Singh is currently working as an Associate Professor in the Department of Computer Science at Delhi College of Arts and Commerce, University of Delhi, Delhi, India. He received his M.C.A. degree from M. M. M. Engineering College, Gorakhpur, U.P., India and Ph.D. degree in Software Reliability from University of Delhi. He has published more than 60 research papers. His research interests include mining software repositories, empirical software engineering and software reliability engineering. He has co-authored a book “Essentials of Computer Network Database and Internet Technology” from NAROSA Publishing India. He has also received an award from Society for Reliability Engineering, Quality and Operations Management for contribution as a promising author and Researcher in the field of computer Science. He has completed research projects from University Grants Commission, Department of Science and Technology Government of India and Uni-versity of Delhi. He is member of Society for Reliability Engineering, Quality and Operations Management and Association of Computing Machinery (ACM).